\documentclass[aps,twocolumn,showpacs]{revtex4}
\usepackage{amsmath}
\usepackage{epsfig}

\begin{document}

\title{Prediction of $N\Omega$-like dibaryons with heavy quarks}

\author{Hongxia Huang$^1$}\email{hxhuang@njnu.edu.cn}
\author{Jialun Ping$^1$}\email{jlping@njnu.edu.cn(Corresponding author)} 
\author{Fan Wang$^2$}\email{284550721@qq.com}

\affiliation{$^1$Department of Physics, Nanjing Normal University,
Nanjing 210097, P.R. China}

\affiliation{$^2$Department of Physics, Nanjing University,
Nanjing 210093, P.R. China}

\begin{abstract}
Possible $N\Omega$-like dibaryons $N\Omega_{ccc}$ and $N\Omega_{bbb}$ with quantum numbers $IJ^P=\frac{1}{2}2^+$
are investigated within the framework of quark delocalization color screening model. We find both of these two 
states are bound, and the binding energy increases as the quarks of the system become heavier. The attraction 
between $N$ and $\Omega_{ccc}$ (or $\Omega_{bbb}$) mainly comes from the kinetic energy term due to quark 
delocalization and color screening. The effect of the channel-coupling provides more effective attraction to 
$N\Omega_{ccc}$ and $N\Omega_{bbb}$ systems.
Besides, the scattering length, the effective range, and the binding energy, obtained from the calculation of the 
low-energy scattering phase shifts, also supports the existence of the $N\Omega_{ccc}$ and $N\Omega_{bbb}$ states. 
All these properties can provide necessary information for experimental search for the $N\Omega$-like dibaryons with
heavy quarks. And the experimental progress can also check the mechanism of the intermediate-range attraction of the
baryon-baryon interaction in quark models.
\end{abstract}


\maketitle

\setcounter{totalnumber}{5}

\section{\label{sec:introduction}Introduction}
The existence of dibaryons is one of the long standing problems in hadron physics. Although the dibaryon searches
experienced several ups and downs in their long history, they received renewed interest in recent years. For the
nonstrange dibaryon, the best-known candidate is the $\Delta\Delta$ resonance state, which was predicted by Dyson 
and Xuong in 1964~\cite{Dyson} and later also by Goldman $et~al.$, who called it the "inevitable dibaryon" $d^{*}$ 
due to its unique symmetry features~\cite{dstar}. Recently, the WASA-at-COSY collaboration reported the discovery 
of this $\Delta\Delta$ resonance state with $M = 2.37$ GeV, $\Gamma \approx 70$ MeV, and $IJ^{P} = 
03^{+}$~\cite{ABC1,ABC2,ABC3}. The detail of this dibaryon observation can be found in Ref.~\cite{Clement}. 
The quark model calculations~\cite{Ping1,Huang1,YuanXQ,HuangF}, as well as the relativistic three-body 
calculations~\cite{Gal1,Gal2} all described properly the characteristics of this resonance.

For the strange dibaryon, the $H$-dibaryon with $J =0,~S=-2$, the $N\Omega$ with $J =2,~S=-3$, and the 
$\Omega\Omega$ with $J =0,~S=-6$ are particularly interesting, since the Pauli blocking among valence quarks 
do not operate in these systems. Above all, the progress of the $N\Omega$ searches in experiment attracted 
more and more attention for its accessible. Very recently, the measurement of the $p\Omega$ correlation 
function was conducted in $Au+Au$ collisions by the STAR experiment at the Relativistic Heavy-Ion Collider 
(RHIC)~\cite{RHIC}, and the result indicated that the scattering length is positive for the $p\Omega$ 
interaction and favored the $p\Omega$ bound state hypothesis. On the theoretical side, the $N\Omega$ state 
has been investigated by several groups. Goldman {\em et al.} predicted that the $S=-3$, $I=1/2$, $J=2$ 
dibaryon state $N\Omega$ might be a narrow resonance in a relativistic quark model~\cite{PRL59}. Oka proposed 
that there should be a quasi-bound state with $IJ^{P}=\frac{1}{2}2^{+}$ by using a constituent quark 
model~\cite{Oka}. Recent study of ($2+1$)-flavor lattice quantum chromodynamics (QCD) simulations by 
HAL QCD Collaboration reported
that the $N\Omega$ was indeed a bound state at pion mass of $875$ MeV~\cite{HAL1} and later with nearly 
physical quark masses ($m_{\pi}\simeq 146$ MeV and $m_{K}\simeq 525$ MeV)~\cite{HAL2}. K. Morita {\em et al.} 
studied the two-pair momentum correlation functions of the dibaryon candidate $N\Omega$ in relativistic 
heavy-ion collisions by employing the interactions obtained from the ($2+1$)-flavor lattice QCD 
simulations~\cite{Morita1,Morita2}. Besides, this state has also been observed to be bound in several 
relativistic quark models~\cite{PRC51,PangHR,ChenM,Huang2,LiQB}.

For the dibaryons with heavy quarks, the $N\Lambda_{c}$ system and the $H$-like dibaryon state
$\Lambda_{c}\Lambda_{c}$ were both studied on hadron level~\cite{Liu1,Liu2} and on quark 
level~\cite{Huang3,Huang4}. The possibility of existing deuteron-like dibaryons with heavy quarks, 
such as $N\Sigma_{c}$, $N\Xi^{\prime}_{c}$, $N\Xi_{cc}$, $\Xi\Xi_{cc}$ and so on, were investigated by 
several realistic phenomenological nucleon-nucleon interaction models~\cite{Fromel,Julia}. Recently 
many near-threshold charmonium-like states called ``$XYZ$" particles were observed, triggering lots 
of studies on the molecule-like bound states containing heavy quarks. Such studies will give further 
information on the hadron-hadron interactions. In the heavy-quark sector, the large masses of the 
heavy quarks reduce the kinetic energy of the system, which makes them easier to form bound states. 
Very recently, the Lattice QCD also studied the deuteron-like dibaryons with valence quark contents:
$\Sigma_{c}\Xi_{cc}(uucucc)$, $\Omega_{c}\Omega_{cc}(sscscc)$, $\Sigma_{b}\Xi_{bb}(uububb)$, $\Omega_{b}\Omega_{bb}(ssbsbb)$, and $\Omega_{ccb}\Omega_{cbb}(ccbcbb)$, and with spin parity $J^{P}=1^{+}$~\cite{Lattice1}. They also found that the binding of these dibaryons became stronger 
as they became heavier in mass. Therefore, the dibaryons with heavy quarks are also possible multiquark 
states, and the study of such system will help us to understand the hadron-hadron interactions and 
search for exotic quark states in temporary hadron physics.

It is known to all that QCD is the fundamental theory of the strong interaction. However, in the 
low-energy region of strong interaction, it is difficult to directly use QCD to study the complicated 
systems such as hadron-hadron interactions and multiquark states because of the nonperturbative 
complication. Various QCD-inspired models have been developed to get physical insights into the 
multiquark systems. The quark delocalization color screening model (QDCSM), developed in the 1990s 
with the aim of explaining the similarities between nuclear and molecular forces~\cite{QDCSM0}, is 
the representative of the quark models. In this model, quarks confined in one nucleon are allowed 
to delocalize to a nearby baryon and the confinement interaction between quarks in different 
baryon orbits is modified to include a color screening factor. The latter is a model description 
of the hidden color channel coupling effect~\cite{Huang5}. The delocalization parameter is determined 
by the dynamics of the interacting quark system, this allows the quark system to choose the most 
favorable configuration through its own dynamics in a larger Hilbert space. The model gives a good 
description of $NN$ and $YN$ interactions and the properties of deuteron~\cite{QDCSM1,QDCSM2}. 
It is also employed to study the dibaryon candidates: $d^{*}$~\cite{Ping1,Huang1} and 
$N\Omega$~\cite{PRC51,PangHR,ChenM,Huang2}, and dibaryons with heavy quarks: the $N\Lambda_{c}$ system 
and the $\Lambda_{c}\Lambda_{c}$ system~\cite{Huang3,Huang4}.

In this work, we continue to investigate dibaryons with heavy quarks by using QDCSM. In our previous 
work, we have shown $N\Omega$ is a narrow resonance in $\Lambda\Xi$ $D$-wave scattering 
process~\cite{ChenM}. However, the $\Lambda$-$\Xi$ scattering data analysis is quite complicated
experimentally. Then we calculated $N\Omega$ scattering length, effective range and binding energy 
based on the low-energy scattering phase shifts. These information can be observed by the 
$N$-$\Omega$ correlation analysis with RHIC and LHC data, or by the new developed automatic scanning 
system at J-PARC~\cite{Huang2}. It is interesting to extend such study to the dibaryons with heavy quarks. 
So we investigate whether the $N\Omega$-like dibaryons: $N\Omega_{ccc}$ and $N\Omega_{bbb}$ exist or 
not in QDCSM. The low-energy scattering phase shifts, scattering length, effective range and binding 
energy of $N\Omega_{ccc}$ and $N\Omega_{bbb}$ are calculated, which are useful for the experimental 
search of such heavy dibaryons.

The structure of this paper is as follows. A brief introduction of QDCSM is given in section II. 
Section III devotes to the numerical results and discussions. The summary is shown in the last section.

\section{The quark delocalization color screening
model (QDCSM)}
The detail of QDCSM used in the present work can be found in the references~\cite{QDCSM0,QDCSM1}. 
Here, we just present the salient features of the model. The model Hamiltonian is:
\begin{widetext}
\begin{eqnarray}
H &=& \sum_{i=1}^6 \left(m_i+\frac{p_i^2}{2m_i}\right) -T_c
+\sum_{i<j} \left[ V^{G}(r_{ij})+V^{\chi}(r_{ij})+V^{C}(r_{ij})
\right],
 \nonumber \\
V^{G}(r_{ij})&=& \frac{1}{4}\alpha_{s} {\mathbf \lambda}_i \cdot
{\mathbf \lambda}_j
\left[\frac{1}{r_{ij}}-\frac{\pi}{2}\left(\frac{1}{m_{i}^{2}}+\frac{1}{m_{j}^{2}}+\frac{4{\mathbf
\sigma}_i\cdot {\mathbf\sigma}_j}{3m_{i}m_{j}}
 \right)
\delta(r_{ij})-\frac{3}{4m_{i}m_{j}r^3_{ij}}S_{ij}\right],
\nonumber \\
V^{\chi}(r_{ij})&=& \frac{1}{3}\alpha_{ch}
\frac{\Lambda^2}{\Lambda^2-m_{\chi}^2}m_\chi \left\{ \left[
Y(m_\chi r_{ij})- \frac{\Lambda^3}{m_{\chi}^3}Y(\Lambda r_{ij})
\right]
{\mathbf \sigma}_i \cdot{\mathbf \sigma}_j \right.\nonumber \\
&& \left. +\left[ H(m_\chi r_{ij})-\frac{\Lambda^3}{m_\chi^3}
H(\Lambda r_{ij})\right] S_{ij} \right\} {\mathbf F}_i \cdot
{\mathbf F}_j, ~~~\chi=\pi,K,\eta \\
V^{C}(r_{ij})&=& -a_c {\mathbf \lambda}_i \cdot {\mathbf
\lambda}_j [f(r_{ij})+V_0], \nonumber
\\
 f(r_{ij}) & = &  \left\{ \begin{array}{ll}
 r_{ij}^2 &
 \qquad \mbox{if }i,j\mbox{ occur in the same baryon orbit} \\
  \frac{1 - e^{-\mu_{ij} r_{ij}^2} }{\mu_{ij}} & \qquad
 \mbox{if }i,j\mbox{ occur in different baryon orbits} \\
 \end{array} \right.
\nonumber \\
S_{ij} & = &  \frac{{\mathbf (\sigma}_i \cdot {\mathbf r}_{ij})
({\mathbf \sigma}_j \cdot {\mathbf
r}_{ij})}{r_{ij}^2}-\frac{1}{3}~{\mathbf \sigma}_i \cdot {\mathbf
\sigma}_j. \nonumber
\end{eqnarray}
\end{widetext}
Where $S_{ij}$ is quark tensor operator; $Y(x)$ and $H(x)$ are standard Yukawa functions; 
$T_c$ is the kinetic energy of the center of mass; $\alpha_{ch}$ is the chiral coupling
constant, determined as usual from the $\pi$-nucleon coupling constant; $\alpha_{s}$ is 
the quark-gluon coupling constant. In order to cover the wide energy range from light, 
strange, to heavy quarks, an effective scale-dependent quark-gluon coupling $\alpha_{s}(u)$ 
was introduced~\cite{Vijande}:
\begin{eqnarray}
\alpha_{s}(u) & = &
\frac{\alpha_{0}}{\ln(\frac{u^2+u_{0}^2}{\Lambda_{0}^2})}.
\end{eqnarray}
The other symbols in the above expressions have their usual meanings. All parameters, which are 
fixed by fitting the masses of baryons with light flavors and heavy flavors, are taken from our 
previous work~\cite{Huang4}. The values of those parameters are listed in Table~\ref{parameters}.
The masses of light flavor baryons are shown in our former work~\cite{Huang2}, here we list the 
masses of the charmed and bottom baryons in Table~\ref{baryons}.

\begin{table}[ht]
\caption{Model parameters:
$m_{\pi}=0.7~{\rm fm}^{-1}$, $m_{K}=2.51~{\rm fm}^{-1}$,
$m_{\eta}=2.77~{\rm fm}^{-1}$, $\Lambda_{\pi}=4.2~{\rm fm}^{-1}$,
$\Lambda_{K}=5.2~{\rm fm}^{-1}$, $\Lambda_{\eta}=5.2~{\rm
fm}^{-1}$, $\alpha_{ch}=0.027$.}
\begin{tabular}{ccccc} \hline
~~~~~~$b$~~~~ & ~~~~$m_{u,d}$~~~~ & ~~~~$m_{s}$~~~~ & ~~~~$m_{c}$~~~~ & ~~~~$m_{b}$~~~~   \\
(fm) & (MeV) & (MeV) & (MeV) & (MeV)   \\ \hline\noalign{\smallskip}
0.518  & 313  &  573 & 1788 &    5141   \\ \hline\noalign{\smallskip}
 $ a_c$ & $V_{0}$ &  $\alpha_{0}$ &  $\Lambda_{0}$ & $u_{0}$   \\
 (MeV\,fm$^{-2}$) & (MeV) &   &  (fm$^{-1}$) & (MeV)   \\\hline\noalign{\smallskip}
   58.03 & -1.2883 &  0.5101 &  1.525 &   445.808   \\
\hline
\end{tabular}
\label{parameters}
\end{table}

\begin{table}[ht]
\caption{The Masses (in MeV) of the charmed and bottom baryons. Experimental values are taken
from the Particle Data Group (PDG)~\cite{PDG}.}
\begin{tabular}{lccccccccc}
\hline
 & ~$\Sigma_{c}$~ & ~$\Sigma^{*}_{c}$~ & ~$\Lambda_{c}$~  &
  ~$\Xi_{c}$~ & ~$\Xi^{*}_{c}$~ & ~$\Xi_{cc}$~ & ~$\Xi^{*}_{cc}$~ & ~$\Omega_{c}$~ & ~$\Omega_{ccc}$~  \\
\hline
 Expt.& 2455 & 2520 & 2286  & 2467 & 2645 & 3621 & $\cdots$ & 2695 & $\cdots$  \\
Model & 2465 & 2489 & 2286 & 2551 & 2638 & 3766 & 3792 & 2786 & 5135  \\
\hline & ~$\Sigma_{b}$~ & ~$\Sigma^{*}_{b}$~ &
~$\Lambda_{b}$~ & ~$\Xi_{b}$~ & ~$\Xi^{*}_{b}$~ & ~$\Xi_{bb}$~ & ~$\Xi^{*}_{bb}$~ &  ~$\Omega_{b}$~ &  ~$\Omega_{bbb}$~   \\
\hline
 Expt.& 5811 & 5832 & 5619 & 5792 & 5955 & $\cdots$ & $\cdots$ & 6046 & $\cdots$  \\
Model & 5809 & 5817 & 5619 & 5888 & 5971 & 10455 & 10464 & 6131 & 15169 \\
\hline
\end{tabular}
\label{baryons}
\end{table}

The quark delocalization in QDCSM is realized by specifying the single particle orbital 
wave function of QDCSM as a linear combination of left and right Gaussians, the single 
particle orbital wave functions used in the ordinary quark cluster model,
\begin{eqnarray}
\psi_{\alpha}(\mathbf{s}_i ,\epsilon) & = & \left(
\phi_{\alpha}(\mathbf{s}_i)
+ \epsilon \phi_{\alpha}(-\mathbf{s}_i)\right) /N(\epsilon), \nonumber \\
\psi_{\beta}(-\mathbf{s}_i ,\epsilon) & = &
\left(\phi_{\beta}(-\mathbf{s}_i)
+ \epsilon \phi_{\beta}(\mathbf{s}_i)\right) /N(\epsilon), \nonumber \\
N(\epsilon) & = & \sqrt{1+\epsilon^2+2\epsilon e^{-s_i^2/4b^2}}. \label{1q} \\
\phi_{\alpha}(\mathbf{s}_i) & = & \left( \frac{1}{\pi b^2}
\right)^{3/4}
   e^{-\frac{1}{2b^2} (\mathbf{r}_{\alpha} - \mathbf{s}_i/2)^2} \nonumber \\
\phi_{\beta}(-\mathbf{s}_i) & = & \left( \frac{1}{\pi b^2}
\right)^{3/4}
   e^{-\frac{1}{2b^2} (\mathbf{r}_{\beta} + \mathbf{s}_i/2)^2}. \nonumber
\end{eqnarray}
Here $\mathbf{s}_i$, $i=1,2,...,n$ are the generating coordinates, which are introduced 
to expand the relative motion wave function. The delocalization parameter 
$\epsilon(\mathbf{s}_i)$ is determined by the dynamics of the quark system rather than
adjusted parameters. It has been used to explain the cross-over transition between hadron 
phase and quark-gluon plasma phase~\cite{Xu}.

\section{The results and discussions}
In this work, we investigate the $N\Omega$-like dibaryons with heavy quarks: $uudccc$ and 
$uudbbb$ systems with quantum numbers $IJ^{P}=\frac{1}{2}2^{+}$ in QDCSM.
The channel coupling effects are also considered. The labels of all coupled channels are 
listed in Table~\ref{channels}.

\begin{center}
\begin{table}[h]
\caption{The channels of $uudccc$ and $uudbbb$ systems.}
\begin{tabular}{lcccccccc}
\hline
 Channels & 1 & ~~~2 & ~~~3 & ~~~4 & ~~~5   \\ \hline
 $J^{P}=2^{+}$~~~ & $\Sigma_{c}\Xi^{*}_{cc}$ & ~~~$\Sigma^{*}_{c}\Xi_{cc}$
                  & ~~~$\Lambda_{c}\Xi^{*}_{cc}$ & ~~~$\Sigma^{*}_{c}\Xi^{*}_{cc}$
                  & ~~~$N\Omega_{ccc}$    \\
 $J^{P}=2^{+}$~~~ & $\Sigma_{b}\Xi^{*}_{bb}$ & ~~~$\Sigma^{*}_{b}\Xi_{bb}$
                  & ~~~$\Lambda_{b}\Xi^{*}_{bb}$ & ~~~$\Sigma^{*}_{b}\Xi^{*}_{bb}$
                  & ~~~$N\Omega_{bbb}$  \\  \hline
\end{tabular}
\label{channels}
\end{table}
\end{center}

In order to check whether or not there is any bound state, a dynamical calculation is needed. 
The resonating group method (RGM)~\cite{RGM} is employed here. By expanding the relative 
motion wave function between two clusters in the RGM equation by gaussians, the 
integro-differential equation of RGM can be reduced to an algebraic equation, which is the 
generalized eigen-equation. Then by solving the eigen-equation, the energy of the system can 
be obtained. Besides, to keep the matrix dimension manageably small, the baryon-baryon 
separation is taken to be less than $6$ fm in the calculation. The binding energies of 
every channel and the channel-coupling of the $uudccc$ and $uudbbb$ systems are listed in 
Table~\ref{bound}, where $B$ stands for the binding energy, $c.c.$ means the result of 
channel-coupling calculation, and $ub$ means that the state is unbound.

\begin{table}[ht]
\caption{The binding energies of the $uudccc$ and $uudbbb$ systems.}
\begin{tabular}{lcccccccc}
\hline
  Channels & ~$\Sigma_{c}\Xi^{*}_{cc}$~ & ~$\Sigma^{*}_{c}\Xi_{cc}$~
                  & ~$\Lambda_{c}\Xi^{*}_{cc}$~ & ~$\Sigma^{*}_{c}\Xi^{*}_{cc}$~
                  & ~$N\Omega_{ccc}$~  & ~$c.c.$~  \\
  $B$ (MeV) & $-1.1$ & $ub$ & $ub$ & $ub$ & $-0.6$ & $-30.9$ \\ \hline
  Channels & ~$\Sigma_{b}\Xi^{*}_{bb}$~ & ~$\Sigma^{*}_{b}\Xi_{bb}$~
                  & ~$\Lambda_{b}\Xi^{*}_{bb}$~ & ~$\Sigma^{*}_{b}\Xi^{*}_{bb}$~
                  & ~$N\Omega_{bbb}$~ & $c.c.$~ \\
  $B$ (MeV) & $ub$ & $ub$ & $ub$ & $ub$ & $-2.8$ & $-50.7$ \\\hline
\end{tabular}
\label{bound}
\end{table}

For the $uudccc$ system, the single channel calculation shows that the both 
$\Sigma_{c}\Xi^{*}_{cc}$ and $N\Omega_{ccc}$ is bound with very small binding energies.
By doing a channel-coupling calculation, the lowest energy of the system is $30.9$ MeV 
lower than the threshold of $N\Omega_{ccc}$, which means that this heavy quark dibaryon 
$N\Omega_{ccc}$ is a bound state in QDCSM. Obviously, the effect of the channel-coupling 
is important for providing more effective attraction to the $N\Omega_{ccc}$ systems.
Besides, the bound state $\Sigma_{c}\Xi^{*}_{cc}$ disappears, because the channel-coupling 
calculation pushes the energy of this state above its threshold. 
For the $N\Omega_{bbb}$ system, the result is similar to the $N\Omega_{ccc}$ system. 
The individual $N\Omega_{bbb}$ channel is bound with binding energy $-2.8$ MeV. 
The channel-coupling calculation lower the energy of the system further, whose energy is 
$50.7$ MeV below the threshold of $N\Omega_{bbb}$, is obtained. By comparing with the 
dibaryon $N\Omega$~\cite{Huang2}, we find that the binding energy increases as the 
quark of the system becomes heavier. This conclusion is consistent with the recent work 
of lattice QCD~\cite{Lattice1}, in which they studied the deuteron-like dibaryons with 
heavy quarks and found that the stability of dibaryons increases as they become heavier.

To investigate the interaction between $N$ and $\Omega_{ccc}$ (or $\Omega_{bbb}$), we calculate 
the effective potentials, $V_{\mbox{\scriptsize{eff}}}=E(s)-E(s=\infty)$, $E(s)$ is the energy of the system
with $N-\Omega_{ccc}$ separation $s$, as well as the contribution of each interaction term to the energy of
the system. The results for $N\Omega_{ccc}$ and $N\Omega_{bbb}$ states are similar. To save space, 
we take the effective potentials between $N$ and $\Omega_{ccc}$ as an example. The total effective 
potentials and the contributions of all interaction terms to the effective potential, including the 
kinetic energy ($V_{vk}$), the confinement ($V_{con}$), the one-gluon-exchange ($V_{oge}$), 
the $\pi$-exchange ($V_{\pi}$), and the $\eta$-exchange ($V_{\eta}$), are shown in Fig.~1. 
We notice that due to the special quark content of $N\Omega_{ccc}$ system, the effective interaction 
have very small contribution from the one-gluon-exchange interaction. The attraction between $N$ and 
$\Omega_{ccc}$ mainly comes from the kinetic energy term due to the quark delocalization, other terms 
provide repulsive potentials, which reduce the total attraction of the $N\Omega_{ccc}$ potential. 
By comparing with the interaction between $N$ and $\Omega$, the behavior is similar, because we use 
the same model, and the mechanism of the intermediate-range attraction in this model is the quark 
delocalization and color screening, which work together to provide short-range repulsion and 
intermediate-range attraction.

\begin{figure}
\epsfxsize=3.0in \epsfbox{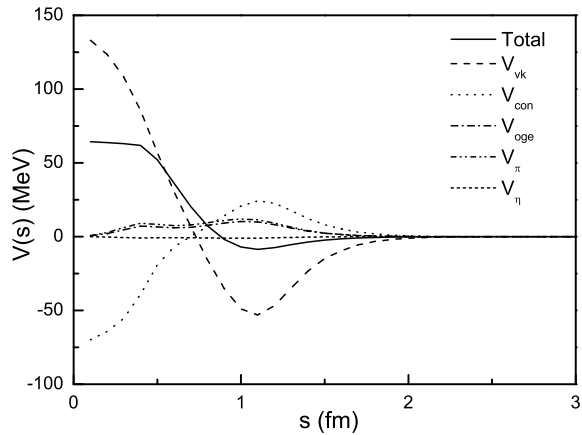} \vspace{-0.1in}

\caption{The contributions to the effective potential from various
terms of interactions.}
\end{figure}

In our previous work, we also calculated the low-energy scattering phase shifts, scattering 
length, and effective range of $N\Omega$, which can provide necessary information for the 
$N$-$\Omega$ correlation analysis with RHIC and LHC data. Naturally, we do the same calculation 
for the $N\Omega_{ccc}$ and $N\Omega_{bbb}$ systems. The low-energy scattering phase shifts 
are calculated by using the well developed Kohn-Hulthen-Kato(KHK) variational method~\cite{RGM}. 
The wave function of the dibaryon system is of the form
\begin{equation}
\Psi = {\cal A } \left[\hat{\phi}_{A}(\boldsymbol{\xi}_{1},\boldsymbol{\xi}_{2})
       \hat{\phi}_{B}(\boldsymbol{\xi}_{3},\boldsymbol{\xi}_{4})\chi_{L}(\boldsymbol{R}_{AB})\right].
\end{equation}
where $\boldsymbol{\xi}_{1}$ and $\boldsymbol{\xi}_{2}$ are the internal coordinates for the baryon
cluster A, and $\boldsymbol{\xi}_{3}$ and $\boldsymbol{\xi}_{4}$ are the internal coordinates for 
another baryon cluster B. $\boldsymbol{R}_{AB} = \boldsymbol{R}_{A}-\boldsymbol{R}_{B}$ is the 
relative coordinate between the two clusters. The symbol ${\cal A }$ is the anti-symmetrization 
operator. The $\hat{\phi}_{A}$ and $\hat{\phi}_{B}$ are the internal cluster wave functions of
two baryons A and B, and $\chi_{L}(\boldsymbol{R}_{AB})$ is the relative motion wave function 
between two clusters. For a scattering problem, $\chi_{L}(\boldsymbol{R}_{AB})$ is expanded as
\begin{equation}
\chi_{L}(\boldsymbol{R}_{AB}) = \sum_{i=1}^{n} C_{i}
    \frac{\tilde{u}_{L}(R_{AB},S_{i})}{R_{AB}}Y_{LM}(\hat{\boldsymbol{R}}_{AB}) .
\end{equation}
with
\begin{eqnarray}
 & & \tilde{u}_{L}(R_{AB},S_{i}) =    \nonumber \\
 & & \left\{ \begin{array}{ll}
 \alpha_{i} u_{L}(R_{AB},S_{i}),   ~~~~~~~~~~~~~~~~~~~~~~~~~~~~~~~~~~R_{AB}\leq R_{C}  \\
  \left[h^{-}_{L}(k_{AB},R_{AB})-s_{i}h^{+}_{L}(k_{AB},R_{AB})\right] R_{AB},  R_{AB}\geq R_{C}
 \end{array} \right. \nonumber \\
\end{eqnarray}
where
\begin{eqnarray}
& & u_{L}(R_{AB},S_{i}) = \sqrt{4\pi}(\frac{3}{2\pi b^2})^{3/4}R_{AB}   \nonumber \\
&& ~~~~\times \exp\left[-\frac{3}{4b^2}(R^{2}_{AB}-S^{2}_{i})\right] i^{L} j_{L}(-i \frac{3}{2b^{2}}R_{AB}S_{i}).
~~~~~
\end{eqnarray}
$\boldsymbol{S}_{i}$ is called the generating coordinate, $C_{i}$ is expansion coefficients,
$n$ is the number of the gaussians, which is determined by the stability of the results.
$h^{\pm}_{L}$ is the $L$-th spherical Hankel functions, $k_{AB}$ is the momentum of relative 
motion with $k_{AB}=\sqrt{2\mu_{AB}E_{cm}}$, $\mu_{AB}$ is the reduced mass of two hadrons 
(A and B) of the open channel, $E_{cm}$ is the incident energy, and $R_{C}$ is a cutoff radius 
beyond which all the strong interaction can be disregarded. $\alpha_{i}$ and $s_{i}$ are 
complex parameters which are determined by the smoothness condition at $R_{AB}=R_{C}$ and 
$C_{i}$ satisfy $\sum_{i=1}^{n} C_{i}=1$. $j_{L}$ is the $L$-th spherical Bessel function.
After performing variational procedure, a $L$-th partial-wave equation for the scattering 
problem can be deduced as
\begin{eqnarray}
& & \sum_{j=1}^{n} {\cal L }^{L}_{ij} C_{j} = {\cal M }^{L}_{i}  ~~~~(i=0,1,\cdot\cdot\cdot, n-1), \label{Lij}
~~~~~
\end{eqnarray}
with
\begin{eqnarray}
& & {\cal L }^{L}_{ij} = {\cal K }^{L}_{ij}-{\cal K }^{L}_{i0}-{\cal K }^{L}_{0j}+{\cal K }^{L}_{00},
~~~~~
\end{eqnarray}
\begin{eqnarray}
& & {\cal M }^{L}_{i}  = {\cal K }^{L}_{00}-{\cal K }^{L}_{i0},
~~~~~
\end{eqnarray}
and
\begin{eqnarray}
& & {\cal K }^{L}_{ij}=\left\langle \hat{\phi}_{A}(\boldsymbol{\xi}'_{1},\boldsymbol{\xi}'_{2})\hat{\phi}_{B}(\boldsymbol{\xi}'_{3},\boldsymbol{\xi}'_{4})
  \frac{\tilde{u}_{L}(R'_{AB},S_{i})}{R'_{AB}}Y_{LM}(\hat{\boldsymbol{R}}'_{AB}) \right.  \nonumber \\
& & ~~~~~~~~~\left|H-E\right| \nonumber \\
& & \left. {\cal A } \left[\hat{\phi}_{A}(\boldsymbol{\xi}_{1},\boldsymbol{\xi}_{2})\hat{\phi}_{B}(\boldsymbol{\xi}_{3},\boldsymbol{\xi}_{4})
  \frac{\tilde{u}_{L}(R_{AB},S_{j})}{R_{AB}}Y_{LM}(\hat{\boldsymbol{R}}_{AB})\right]\right\rangle. \nonumber \\
\end{eqnarray}
By solving Eq.(\ref{Lij}), we can obtain the expansion coefficients $C_{i}$. Then the 
scattering matrix element $S_{L}$ and the phase shifts $\delta_{L}$ are given by
\begin{eqnarray}
& & S_{L} \equiv e^{2i\delta_{L}} = \sum_{i=1}^{n} C_{i}s_{i},
~~~~~
\end{eqnarray}
Then, we can extract the scattering length $a_{0}$ and the effective range $r_{0}$ from 
the low-energy phase shifts by using the formula:
\begin{eqnarray}
k_{AB}cot\delta_{L} & = & -\frac{1}{a_{0}}+\frac{1}{2}r_{0}k_{AB}^{2}+{\cal
O}(k_{AB}^{4})
\end{eqnarray}
Finally, the binding energy $B^{\prime}$ is calculated according to the relation:
\begin{eqnarray}
B^{\prime} & = &\frac{\hbar^2\alpha^2}{2\mu_{AB}}
\end{eqnarray}
where $\alpha$ is the wave number which can be obtained
from the relation~\cite{Babenko}:
\begin{eqnarray}
r_{0} & = &\frac{2}{\alpha}(1-\frac{1}{\alpha a_{0}})
\end{eqnarray}
Please note that here we use another method to calculate the binding energy, so we label 
it as $B^{\prime}$. The low-energy phase shifts are shown in Fig. 2, and the scattering 
length, the effective range, as well as the binding energy $B^{\prime}$ are listed in 
Table~\ref{length}. All results are obtained with the five channels coupling calculation in QDCSM.

\begin{figure}
\epsfxsize=3.0in \epsfbox{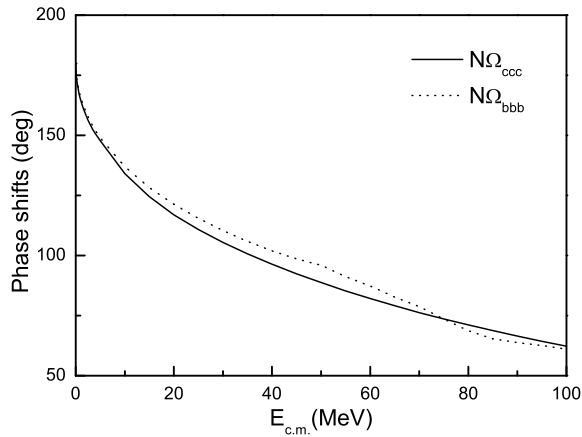} \vspace{-0.1in}

\caption{The phase shifts of both $N\Omega_{ccc}$ and $N\Omega_{bbb}$ dibaryons.}
\end{figure}

It is obvious that the scattering phase shifts of both $N\Omega_{ccc}$ and $N\Omega_{bbb}$ 
states go to $180$ degrees at $E_{c.m.}\sim 0$ and rapidly decreases as $E_{c.m.}$ increases, 
which implies the existence of the bound states. The change rule is similar to the 
$N\Omega$ state~\cite{Huang2}. Besides, the results are consistent with the bound 
state calculation shown above.

\begin{center}
\begin{table}[h]
\caption{The scattering length $a_{0}$, effective range $r_{0}$,
and binding energy $B^{\prime}$ of the $N\Omega_{ccc}$ and $N\Omega_{bbb}$ dibaryons.}
\begin{tabular}{lcccc}
\hline
  & ~~$a_{0}$~(fm)~~ & ~~$r_{0}$~(fm)~~ & ~~$B^{\prime}$~(MeV)~~    \\ \hline
 ~~~$N\Omega_{ccc}$~~~ & 1.3347  & 0.43343 & -21.6    \\
 ~~~$N\Omega_{bbb}$~~~ & 1.1608  & 0.53617 & -40.1   \\
  \hline
\end{tabular}
\label{length}
\end{table}
\end{center}

From Table~\ref{length}, we can see that the scattering length of both $N\Omega_{ccc}$ 
and $N\Omega_{bbb}$ states are all positive, which implies again that these two states 
are bound states. The binding energies $B^{\prime}$ of these two states are close to the 
binding energy $B$ shown above. It indicates that the binding energy from the two methods 
are coincident with each other.

\section{Summary}
In this work, we investigate the $N\Omega$-like dibaryons with heavy quarks: $uudccc$ and 
$uudbbb$ systems with quantum numbers $IJ^{P}=\frac{1}{2}2^{+}$ in the framework of QDCSM. 
Our results show that both of these two states are bound. In quark model, the hadron-hadron 
interaction usually depends critically upon the contribution of the color-magnetic interaction. 
However, due to the special quark content of $N\Omega_{ccc}$ and $N\Omega_{bbb}$ systems, 
the effective interaction have very small contribution from the color-magnetic interaction. 
The attraction between $N$ and $\Omega_{ccc}$ (or $\Omega_{bbb}$) mainly comes from the kinetic 
energy term due to quark delocalization and color screening. Besides, the channel-coupling also 
plays an important role for providing more effective attraction to the $N\Omega_{ccc}$ and 
$N\Omega_{bbb}$ systems. This rule is similar to the $N\Omega$ system. Searching for the 
$N\Omega$ bound state has made considerable headway by the STAR experiment. If the existence 
of $N\Omega$ state can be confirmed by more experiments, it will be a signal showing that 
the mechanism of quark delocalization and color screening is really responsible for the 
intermediate range attraction of baryon-baryon interaction.

On the other hand, we find that the binding of these dibaryons becomes stronger as they 
become heavier in mass, which indicates that it is more possible for the $N\Omega_{ccc}$ 
and $N\Omega_{bbb}$ states to be bound. So it is worth looking for such $N\Omega$-like 
dibaryons in the experiment, although it will be a challenging subject.
Besides, the low-energy scattering phase shifts, the scattering length, the effective range, 
and the binding energy (obtained from the scattering length) also support the existence of 
the $N\Omega_{ccc}$ and $N\Omega_{bbb}$ states. All these characteristic can provide necessary
information for experimental search for the $N\Omega$-like dibaryons with heavy quarks.

\acknowledgments{This work is supported partly by the National Science Foundation of China under
Contract Nos. 11675080, 11175088 and 11535005.}

\end{document}